\documentclass[prl,nofootinbib,showpacs,showkeys,twocolumn]{revtex4}
\usepackage{exscale}
\usepackage{amssymb,amsbsy}
\usepackage[dvips]{graphicx}
\def\bphi{\boldsymbol\phi}
\def\bgam{\boldsymbol\gamma}

\begin{document}
\title{Attenuation of $\bphi$ mesons in $\bgam\bf A$ reactions}
\author{P.~M\"uhlich}
\author{U.~Mosel}
\affiliation{Institut f\"ur Theoretische Physik, Universit\"at
Giessen, D--35392 Giessen, Germany}
\begin{abstract}
We present a theoretical analysis of inclusive photoproduction of
$\phi$ mesons in nuclei. In particular the dependence of the total
$\phi$ meson yield on the target mass number is investigated. The
calculations are done using the semi-classical BUU transport
approach that combines the initial state interaction of the incoming
photon with the coupled-channel dynamics of the final state
particles. The conditions of the calculations are chosen such as to
match the set up of a recent experiment performed at
SPring8/Osaka. Whereas the observables prove to be rather sensitive
to the $\phi$ self energy in the medium, the attribution of
deviations from the standard scenario to a particular in-medium
effect seems to be impossible.
\end{abstract}
\keywords{photonuclear reactions, $\phi$ decay in nuclei, medium
properties of mesons} \pacs{25.20.Lj, 13.25.-k, 14.40.Cs} \maketitle

The question of how the properties of the light vector mesons change
once they are put into a strongly interacting environment provides a
field of very active research. In particular the theoretical and
experimental determination of the vector and isoscalar spectral
densities in systems with either finite temperature or baryon
density has attracted much attention \cite{Rapp:1999ej}. In
particular the $\phi$ meson provides a well suited probe as it is a
very sharp resonance in vacuum.  One could thus expect to observe
even smallest deviations of its spectral distribution in the nuclear
medium from its properties in vacuum. Numerous authors have studied
the medium modifications of the $\phi$ properties in different
approaches as effective Lagrangians and QCD sum rules
\cite{Klingl:1997kf,Klingl:1997tm,Oset:2000eg,Cabrera:2002hc,Cabrera:2003wb}.
As a general picture from these studies emerges a quite small shift
of the $\phi$ meson pole mass, but a sizeable renormalization of its
width as temperature and/or density increase. Particular numbers for
these changes at nuclear saturation density and vanishing
temperature are for instance a width of 28 MeV and a mass shift of
$-6$ MeV found in the studies of
Refs.~\cite{Cabrera:2002hc,Cabrera:2003wb} and 40 MeV/$-10$ MeV
obtained by the authors of Ref.~\cite{Klingl:1997kf,Klingl:1997tm}.
The origin of the $\phi$ medium modifications has been attributed to
either $\phi$-nucleon and/or $\phi$-meson collisions and a change of
the $K\bar K$ self energy caused by the renormalization of the kaon
and antikaon properties inside the nuclear many-body system.

The most direct approach to observe the $\phi$ in-medium properties
would be a reconstruction of its mass distribution from the dilepton
decay channel since this final state is essentially free of any
final state interactions. $\phi$ mesons have indeed been seen in
such an experiment on cold nuclear matter at JLAB; its evaluation is
presently underway \cite{Tur:2004}. In a hot nuclear environment the
dilepton line has been seen in an ultrarelativistic heavy-ion
experiment at the SPS \cite{Damjanovic:2005ni}. Exploiting the dominant decay
branch of the $\phi$ would be a measurement of the $K\bar K$
invariant mass spectrum. The possibility to study modifications of
the $\phi$ meson in photon-nucleus reactions has first been
considered by the authors of Ref.~\cite{Oset:2000na}. These authors
have proposed to measure the $K^+K^-$ mass spectrum from $\phi$
mesons produced in finite nuclei with momenta smaller than $100-150$
MeV. The condition of small $\phi$ momenta ensures that a large
fraction of events in the data sample stem from $\phi$ decays inside
the nucleus, hence carrying information about the in-medium $\phi$
spectrum to the detectors. However, later we have shown
\cite{Muhlich:2002tu} that such a measurement is not without
problems: First, due to the strongly forward peaked angular
distribution, the photon-nucleus cross section including such a
restrictive momentum cutoff is extremely small. Second, the small
kaon mean free path on the one hand reduces the nuclear densities
probed due to $K^-$ absorption and on the other hand distorts the
$K^+K^-$ invariant mass spectrum due to quasi elastic $K^+N$ and
$K^-N$ scattering processes. Moreover, even the nuclear Coulomb
potential, that is much longer ranged than the strong interaction
phenomenon to be probed, makes it impossible to gather any valuable
information about the in-medium $\phi$ properties. Consequences of
these effects have been investigated in great detail in
Ref.~\cite{Muhlich:2002tu}.

Despite this somewhat discouraging situation the authors of
Ref.~\cite{Cabrera:2003wb} have pointed out that there is a
possibility to study at least the imaginary part of the $\phi$ self energy in nuclei by an
attenuation measurement of the $\phi$ flux in nuclear $\phi$
photoproduction. This is a method that has been used already in early
experiments on the $\rho$ meson properties in nuclei
\cite{Alvensleben:1970uw} and has led there to the first extraction
of the $\rho N$ cross section from photoproduction experiments. For
the $\phi$ meson this suggestion has been taken up by a nuclear
$\phi$ photoproduction experiment at SPring8/Osaka that is optimized
to measure $\phi$ mesons under forward angles by detecting the
$K^+K^-$ decay channel \cite{Ishikawa:2005aw}. The observable in
this experiment is the so-called nuclear transparency ratio
\begin{eqnarray}\label{ta}
T_A=\frac{\sigma_{\gamma A\rightarrow\phi X}}{A\sigma_{\gamma N\rightarrow\phi X}},
\end{eqnarray}
i.~e. the ratio of the nuclear $\phi$ production cross section
divided by $A$ times the cross section on a free nucleon. It can be
interpreted as the momentum and position space averaged probability
of a $\phi$ meson to get out of the nucleus. The dependence of the
loss of flux on the target mass number is related to the
absorptive part of the $\phi$-nucleus potential and thus to the
$\phi$ width in the nuclear medium.

In a simple Glauber approximation, neglecting Fermi motion, Pauli
blocking, coupled-channel effects, nuclear shadowing and quasi
elastic scattering processes, the nuclear cross section for $K^+K^-$
photoproduction via the exclusive incoherent production of $\phi$
mesons can be written as
\begin{eqnarray}\label{glauber}
\sigma_{\gamma A}&=&\int d\Omega\int d^3r\,\rho({\bf r})
\frac{d\sigma_{\gamma N}}{d\Omega} \exp \left[ -\sigma_{\phi
N}^{\mathrm{inel}} \int\limits_0^{|{\mathbf{
\Delta}}|}\,dl\,\rho({\mathbf{r}}')\right]
\nonumber\\
& &\times \,
F^\mathrm{K}_{\mathrm{abs}}({\mathbf{r}}+{\mathbf{\Delta}})
\end{eqnarray}
with
\begin{eqnarray}
{\bf r}' & = & {\bf r}+l\frac{\bf q}{|\bf q|}\qquad\mathrm{and}\\
{\bf \Delta} & = & \frac{v}{\gamma}\frac{1}{\Gamma_{\phi}^{\mathrm{dec}}}
\frac{\bf q}{|\bf q|}.
\end{eqnarray}
$F^{\mathrm{K}}_{\mathrm{abs}}$ is a $K^-$ absorption factor that is
obtained by integrating the $K^-$ absorption probability along the
$K^-$ trajectory from the decay vertex of the $\phi$ at ${\bf
r}+{\bf\Delta}$ to infinity and averaging over the possible $K^-$
directions. The Lorentz factor $\gamma$ transforms the $\phi$ width
from the $\phi$ rest frame to the $\phi$ moving frame. In the limit
of $\Gamma_{\phi}^{\mathrm{dec}}\rightarrow0$ and a vanishing $\phi$
absorption cross section $\sigma_{\phi N}^{\mathrm{inel}}\rightarrow
0$ the exponential as well as the $K^-$ absorption factor become
equal to unity. This also implies a nuclear transparency ratio of
unity as can be seen from the definition Eq.~(\ref{ta}).

Using now the low-density theorem that relates the total $\phi N$
cross section to the $\phi$ collision width and the relation
\begin{eqnarray}
-\frac{\mathrm{Im}\Pi_{\phi}^{\mathrm{}}}{\omega}=\gamma\Gamma_{\phi}^{\mathrm{coll}}=
\rho \sigma v_\phi
\end{eqnarray}
the exponential in Eq.~(\ref{glauber}) can be rewritten as
\begin{eqnarray}\label{expo}
\exp\left[\frac{1}{q}\int\limits_0^{|{\mathbf{
\Delta}}|}\mathrm{Im}\Pi\left(q,\rho({\mathbf{r}}')\right)
dl\right].
\end{eqnarray}
$\Gamma_{\phi}^{\mathrm{coll}}$ is the in-medium $\phi$ collisional
width corresponding to nuclear quasi elastic and absorption
channels. The $K^+K^-$ decay channel is omitted from the self energy
in Eq.~(\ref{expo}) since $\phi$ mesons decaying to this channel
will be detected and, hence, do not lead to a loss of flux.

From this expression one can now read off how two different
mechanisms will affect the nuclear transparency ratio: First, if the
$\phi$ collision width becomes large because of the opening of
inelastic nuclear channels, the nuclear cross section and the
transparency ratio will become smaller because of the exponential
suppression factor. Second, if the $\phi$ decay width to the
$K^+K^-$ channel increases e.g. due to kaon self energies in the
medium, the $\phi$ decay vertex at ${\bf r}+{\bf \Delta}$ lies with
higher probability inside the radius of the target nucleus.
Antikaons produced inside a strongly attractive nuclear potential
will be confined for a longer time to the nuclear volume or may even
be bound. Due to the large imaginary part of the nuclear $K^-$
potential these antikaons will be absorbed to inelastic nuclear
channels.

In the very same spirit as in \cite{Alvensleben:1970uw} for the case
of $\rho$ photoproduction in nuclei the authors of
\cite{Ishikawa:2005aw} used a simple Glauber formula in order to
extract the total $\phi N$ cross section from the measured nuclear
$\phi$ photoproduction cross sections. To this end nuclear effects
as Fermi motion, Pauli blocking, quasi elastic scattering, etc. have
been neglected. A similar approach has been used in
Ref.~\cite{Cabrera:2003wb} but including correction factors in order
to account for the effects of Pauli blocking and Fermi motion.
Moreover, a sophisticated model for the $\phi$ nucleon interaction
based on chiral SU(3) dynamics has been incorporated. Nevertheless
the experimental attenuation effect has been underestimated by
almost a factor of two. In the present paper we aim at a careful
treatment of \emph{all} nuclear effects -- i.~e. Fermi motion, Pauli
blocking, shadowing, elastic and inelastic scattering including
sidefeeding and regeneration of all produced particles, decays of
unstable particles, collisional braodening, kaon self energies -- in
order to verify any modification of the $\phi N$ interaction going
beyond the standard cross sections estimated from vacuum processes.


In order to calculate the production and propagation of $\phi$
mesons in finite nuclei we use the semi-classical BUU transport
theory. This theory describes incoherent photon-nucleus reactions.
It has been applied previously to the study of medium modifications
of the $\rho$, $\omega$ and $\phi$ mesons by means of dilepton
\cite{Effenberger:1999ay}, $\pi^0\gamma$ \cite{Muhlich:2003tj} and
$K^+K^-$ \cite{Muhlich:2002tu} photoproduction in nuclei.  All these
reactions are modeled as a two-step process: In the first step the
incoming (and potentially shadowed) photon interacts with a single
nucleon in the target nucleus and produces one or several particles.
In the second step the products of this elementary interaction are
propagated through the nuclear many-body system by means of the semi-classical
transport equations. During the reaction nuclear effects such as Fermi
motion, Pauli blocking, binding energies and shadowing are taken
into account. Cross sections for the elementary photon-nucleon
interaction are constrained by experimental data as far as possible;
for details on the cross sections used for these primary processes
see \cite{Muhlich:2002tu}. Inclusive processes that have not been
determined experimentally are modeled by means of the Lund model
FRITIOF. During the propagation we include all kinds of final state
interactions such as absorption and elastic and inelastic scattering
of the produced particles from the target nucleons. The cross sections for
these processes are obtained either from vacuum matrix elements that
have been fixed using experimental data or again within the Lund
model. A more detailed presentation of the BUU model can be found in
Refs.~\cite{Teis:1996kx,Lehr:1999zr,Muhlich:2002tu,Effenberger:1999ay,Falter:2002jc}
and references therein.

In order to explore the $\phi$ nucleon interaction in the nuclear
medium we use a very basic model. For the total $\phi N$ cross
section we adopt the parametrization from
Refs.~\cite{Schuler:1995fk,Schuler:1996fc} that has been obtained
using Regge theory and an additive quark model:
\begin{eqnarray}\label{phitot}
\sigma_{\phi p}& \simeq & \sigma_{K^+p}+\sigma_{K^-p}-\sigma_{\pi^+p}\nonumber\\
& \simeq & \left(10.01s^{\epsilon}-1.52s^{-\eta}\right)~\mathrm{mb}
\end{eqnarray}
with $s$ in $\mathrm{GeV}^2$ and the Regge intercept parameters
$\epsilon=0.0808$ and $\eta=0.4525$. Because of isospin symmetry the
cross section on protons and neutrons are the same. We note that
this quark model estimate is in line with the common values of
$8-12$ mb quoted in the literature \cite{Ishikawa:2005aw}.

For the elastic channel one can estimate the total cross
section within the strict vector meson dominance model to amount to roughly 10 mb.
Throughout our transport simulations we use the parameterization from Ref.~\cite{Golubeva:1997na}
\begin{eqnarray}\label{phi_elastic}
\sigma_{\phi N}^{\mathrm{el}}=\frac{10}{1+|\bf q|}~\mathrm{mb}
\end{eqnarray}
with $\bf q$ in GeV. For invariant energies above 2.2 GeV we rely on
the FRITIOF routine to simulate the inelastic $\phi N$ scattering
events. For lower energies we treat the inelasticity
$\sigma^{\mathrm{inel}}=\sigma^{\mathrm{tot}}-\sigma^{\mathrm{el}}$
as being due to $\phi$ meson absorption. The process $\phi
N\rightarrow\pi N$ is constrained by its inverse that has been
fitted to experimental data in Ref.~\cite{Sibirtsev:1996uy}. The
remaining inelasticity we put entirely into the channel $\phi
N\rightarrow 2\pi N$. This is justified because for inclusive
observables as the total nuclear cross section coupled channel
effects involving sidefeeding of the $\phi$ meson channel
are expected to play a minor role.

The collisional part of the self energy of the $\phi$ meson we
obtain via the low density theorem taking now the Fermi-distribution
into account
\begin{eqnarray}\label{gcoll}
\mathrm{Im}\Pi(q_0=\sqrt{m_{\phi}^2+{\bf q}^2},{\bf q})=-4\int\frac{d^3p}{(2\pi)^3}
\Theta(|{\bf p}|-p_F)\nonumber\\ \times\frac{p\sqrt{s}}{E_N({\bf p})}
\sigma_{\phi N}^{\mathrm{tot}}(s)\quad
\end{eqnarray}
with $p$ being the center-of-mass momentum of nucleon and $\phi$,
$E_N$ the nucleon on-shell energy in the laboratory frame and $p_F$
the local Fermi momentum. The cross section $\sigma_{\phi
N}^{\mathrm{tot}}$ is the total $\phi$ nucleon cross section
containing all quasi elastic and absorption channels. For $\phi$
mesons at rest we find a collision width of 18 MeV at normal nuclear
matter density, in agreement with the more refined calculations of
\cite{Cabrera:2002hc,Cabrera:2003wb}.

\begin{figure}
\begin{center}
\includegraphics[scale=.85]{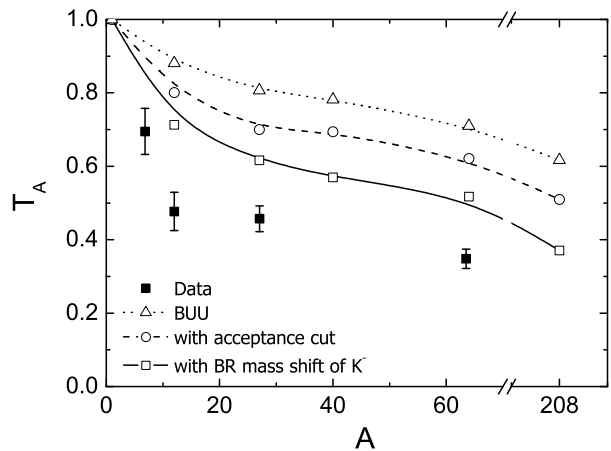}
\caption{Nuclear transparency ratio from BUU simulations (open
symbols) in comparison to the experimental data (solid symbols) from
Ref.~\cite{Ishikawa:2005aw}. The lines are spline interpolations and
are drawn in order to guide the eye.}\label{figure1}
\end{center}
\end{figure}

The total $\phi$ meson yield has been measured at SPring8 from Li,
C, Al and Cu nuclei at photon energies of $E_{\gamma}=1.5-2.4$ GeV.
We start our calculations with including only collisional broadening
of the $\phi$ meson as an in-medium effect while maintaining the
vacuum properties for all other mesons (in particular the kaons).
Results for the transparency ratio are shown in Fig.~\ref{figure1}
by the open triangles. The deviation of the transparency ratio from
unity is mainly due to two effects: First, the $\phi$ meson or its
decay products (if the $\phi$ decays inside the nucleus) are
absorbed on their way out of the nucleus. Second, the transparency
ratio is reduced by Pauli blocking. Due to the diffractive
production mechanism the $\phi$ photoproduction cross section is
peaked in forward direction. Hence the momentum transfer to the hit
nucleon in general is quite small. As a consequence, the momentum of
the final state of the nucleon lies with high probability again
within the Fermi sphere, leading to Pauli blocking of this reaction.
In the energy regime considered here the coherence length of the
$\phi$ component in the incoming photon is still small as compared
to the $\phi$ mean free path. Thus, nuclear shadowing leads only to
a marginal correction of the nuclear cross section.

The results depicted by the open circles show a further reduction of
the transparency ratio due to the limited acceptance of the LEPS
spectrometer that can detect kaons and antikaons only under forward
angles \cite{Nakano:2003qx,Zegers:2003ux}. The nonzero transverse
momentum of the hit target nucleon due to Fermi motion leads to a
broadening of the angular distributions of the produced $\phi$
mesons for finite nuclear targets as compared to the reaction on a
free nucleon at rest that is forward peaked. Hence, more of the
produced mesons do not fall in the acceptance window, leading to the
observed reduction of the nuclear cross section when the geometrical
acceptance constraint is turned on. This reduction is thus not
connected to the absorptive part of the $\phi$ meson selfenergy as
some of the $\phi$ mesons just go into another direction that is not
covered by the forward detector setup.

So far, we have assumed that the branching ratio for the decay of the
$\phi$ meson into $K^+K^-$ is in medium the same as in vacuum.
However, in the nuclear environment kaons and antikaons experience
strong potentials due to the kaon nucleon interaction
\cite{Kaplan:1986yq,Schaffner:1996kv,Waas:1997pe,Lutz:1997wt,Ramos:1999ku,Tolos:2002ud}.
For the antikaon these potentials lead to a considerable
renormalization of its mass whereas for the kaon a cancellation of
scalar and vector potentials results in an at most slight repulsive
mass shift. To explore the effect of these potentials we adopt the
approach of Ref.~\cite{Fuchs:1998yy} where the influence of a
relativistically correct implementation of the nuclear kaon dynamics
on the $K^+$ flow in heavy ion collisions has been investigated.
Relying on the chiral Lagrangian set up by Kaplan and Nelson and
applying the mean field approximation the following kaon dispersion
relation has been obtained:
\begin{eqnarray}\label{disprel}
E_{K^{\pm}}=\sqrt{{m_K^*}^2+({\bf p}\pm{\bf V})^2}\pm V_0
\end{eqnarray}
with the effective kaon mass
\begin{eqnarray}\label{effmass}
m_K^*=\sqrt{m_K^2-\frac{\Sigma_{KN}}{f_{\pi}^2}\rho_s+V_{\mu}V^{\mu}}
\end{eqnarray}
and the vector potential
\begin{eqnarray}
V_{\mu}=\frac{3}{8{f_{\pi}^*}^2}j_{\mu},
\end{eqnarray}
where $j_{\mu}$ is the baryon four vector current. In the case at
hand the spatial components of this current vanish since the target
nucleus stays close to its ground state during the photonuclear
reaction. For the kaon nucleon sigma term we adopt the value of
$\Sigma_{KN}=450$ MeV from the mean field approach of
Ref.~\cite{Brown:1995ta}. Further parameters of the model are the
vacuum pion decay constant $f_{\pi}=93~\mathrm{MeV}$ and the
in-medium pion decay constant at normal nuclear matter density
$f_{\pi}^*=\sqrt{0.6}f_{\pi}$.

The relativistic kaon potentials lead to a considerable reduction of
the in-medium $K^-$ mass whereas the $K^+$ mass increases only
slightly. This leads to a significantly larger phase space for the
$\phi\rightarrow K^+K^-$ decay. The $\phi$ decay width corresponding
to the $K^+K^-$ self energy diagram is proportional to $p^3$, where
$p$ is the kaon/antikaon momentum in the $\phi$ rest frame. At
normal nuclear matter density this width can become as large as 20
MeV (2.18 MeV in vacuum). Note, that taking together the effects of
the kaon renormalization as well as $\phi$ nucleon collisions the
in-medium $\phi$ width can add up to values of about 40 MeV at
saturation density.

\begin{figure}
\begin{center}
\includegraphics[scale=.85]{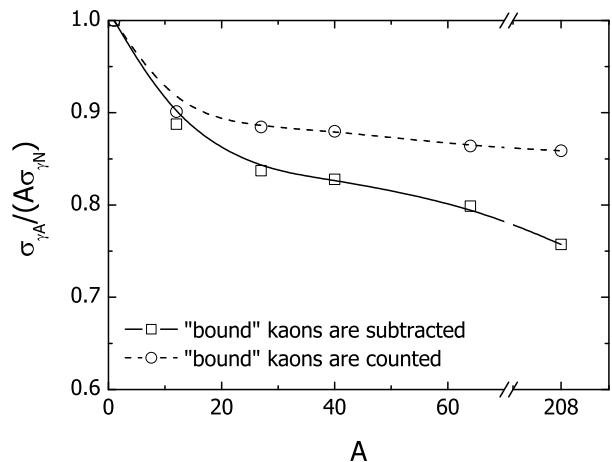}
\caption{Nuclear transparency ratio as function of the target mass
from BUU simulations. No acceptance constraints are
applied.}\label{figure2}
\end{center}
\end{figure}

As a consequence of the larger decay width, more of the produced
$\phi$ mesons decay inside the target nucleus. The antikaons
produced inside the strongly attractive potential can be confined to
the nuclear volume leading to an increase of $K^-$ absorption.
Alternatively, their trajectories are distorted due to the
propagation through the nonzero potential gradients. These $K^+K^-$
pairs are likely to be found outside the experimentally imposed
acceptance window. The results for the nuclear transparency ratio
including the kaon/antikaon dispersion relation from
Eq.~(\ref{disprel}) are shown in Fig.~\ref{figure1} by the open
squares. Indeed the expected reduction is observed. We illustrate
the effect of the antikaon potential in Fig.~\ref{figure2}. In one
curve shown there the antikaons with negative total energy are
subtracted from the data sample, whereas for the second curve also
these not detectable particles are counted to obtain the total
antikaon yield.´

However, even after including all these in-medium effects, the
strength of the experimentally measured attenuation can still not be
reproduced, see Fig.~\ref{figure1}. This discrepancy implies an
additional in-medium effect. A further reduction of the nuclear
cross section could in principle be caused by either a modification
of the $K^+K^-$ decay width going beyond our simple approach, a
modified $\phi N$ absorption cross section or even a change of the
primary production processes. Moreover, the renormalization of the
kaon properties in the medium could also cause a modification of the
$K^+N$ and $K^-N$ cross sections. Such effects have, for instance,
been studied in Ref.~\cite{Schaffner-Bielich:1999cp,Korpa:2004ae}.
However, for the time being we disregard the effects of such
additional medium corrections and consider
the whole attenuation effect as being due to $\phi$ meson
absorption. The comparison to the experimental data then fixes a
value for the total $\phi N$ absorption cross section. Possible changes
of the involved initial and final state processes, as discussed,
introduce considerable ambiguities in the extraction of this
quantity.

\begin{figure}
\begin{center}
\includegraphics[scale=.85]{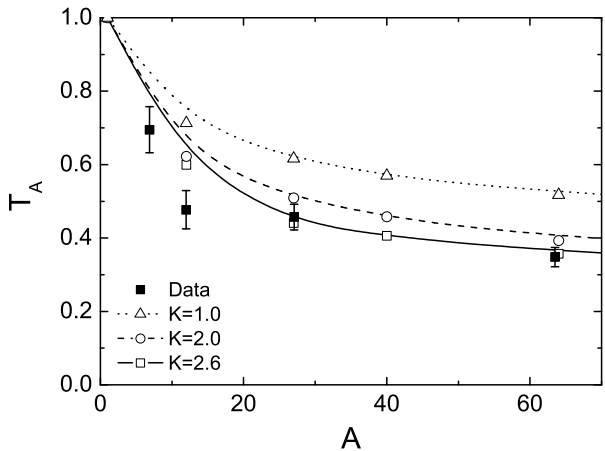}
\caption{Nuclear transparency ratio as function of the target mass.
Shown are BUU calculations (open symbols) in comparison to the data
(solid squares) from Ref.~\cite{Ishikawa:2005aw}. The lines are
spline interpolations of the calculated data points. Calculations
have been done for different $K$-factors, see text.}\label{figure3}
\end{center}
\end{figure}

In order to fit the $\phi N$ cross section to the experimental
transparency data we make the following ansatz: We multiply the
total $\phi N$ cross section given by Eq.~(\ref{phitot}) with a
constant normalization $K$-factor
\begin{eqnarray}
\tilde\sigma_{\phi N}=K\cdot\sigma_{\phi N}.
\end{eqnarray}
In doing so we keep the partial channels $\phi N\rightarrow\phi N$
and $\phi N\rightarrow\pi N$, which are at least roughly constrained
by experimental data, untouched. Thus, the modification of the
$\phi$ in-medium cross section is entirely moved into the absorptive
channel $\phi N\rightarrow 2\pi N$. The results of these
calculations are shown in Fig.~\ref{figure3}. Best agreement is
obtained with a $K$-factor of $K=2.6$. The momentum spectrum of
$\phi$ mesons in the simulated data sample inside the acceptance
window is almost symmetrically distributed around a mean value of
about 1.3 GeV.  This leads to a total $\phi N$ cross section of
$\tilde\sigma_{\phi N}\simeq 27~\mathrm{mb}$ indicated by the data.

\begin{figure}
\begin{center}
\includegraphics[scale=.85]{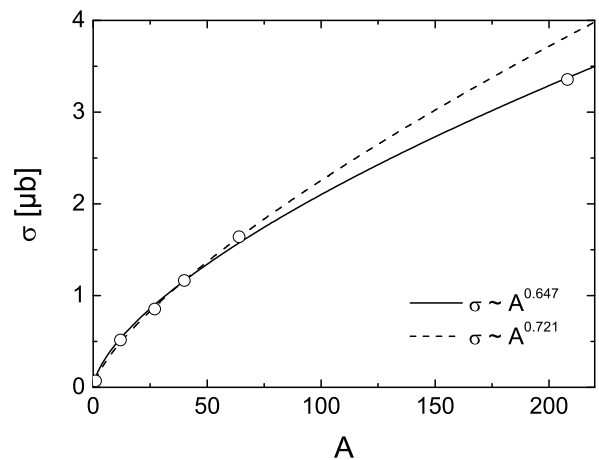}
\caption{$A$-dependence of the total cross section. The open symbols
are the results of our BUU calculations. The solid line is a fit to
all calculations, whereas for the dashed line the $^{208}$Pb nucleus
has been omitted from the fit.}\label{figure4}
\end{center}
\end{figure}

A further quantity extracted from the experimental data in
Ref.~\cite{Ishikawa:2005aw} is the $A$-dependence of the total
$\phi$ meson yield. A scaling close to $\sigma\sim A^{2/3}$ implies
strong absorption as the total cross section scales with the size of
the nuclear surface. On the other hand a scaling close to
$\sigma\sim A$ implies weak absorption as all nucleons of the target
contribute to the total cross section. Considering only the
incoherent events the authors of \cite{Ishikawa:2005aw} have fitted
the total yield with the ansatz
\begin{eqnarray}\label{siga}
\sigma(A)\propto A^{\alpha}
\end{eqnarray}
with a value of $\alpha=0.72\pm0.07$. The nuclear cross section as
function of the target mass number from our BUU calculations that
agree best with the experimental transparency ratio ($K=2.6$) is
shown in Fig.~\ref{figure4}. Fitting the ansatz (\ref{siga}) to all
calculated nuclei we find a value of $\alpha\simeq 0.65$. Omitting
the lead target in the fit, that also has not been considered
experimentally, we obtain the value $\alpha\simeq 0.72$. This is in
perfect agreement with experiment. Anyway, both values clearly show
that the production of $\phi$ mesons on nuclei is surface dominated.

In summary we find that the change of the kaon self-energies in
medium has a significant effect on the observed attenuation of the
$\phi$ meson yield from finite nuclei. However, even with the
inclusion of this in-medium change the measured transparency cannot
be fully explained. In this respect we confirm one of the main
conclusions from \cite{Ishikawa:2005aw,Cabrera:2002hc}. Considering
all additional medium effects as being due to $\phi$ meson
absorption, including a proper relativistic treatment of the
kaon/antikaon potentials in our BUU simulations, we find that a
value of $\sigma_{\phi N}\approx 27~\mathrm{mb}$ is needed to obtain
agreement with the data. This value is considerably higher than
usual quark model estimates for the $\phi N$ cross section in
vacuum. The nuclear transparency ratio as well as the $A$-dependence
of the total $\phi$ meson yield can be reproduced with high accuracy
by adopting this value for the $\phi N$ cross section.

The authors acknowledge stimulating discussions with T.~Falter and
A.~Larionov. We also gratefully acknowledge support by DFG, BMBF and
the Frankfurt Center for Scientific Computing.


\end{document}